\documentclass[journal]{IEEEtran}
\usepackage{cite}
\usepackage{enumitem}
\usepackage{footnote}
\usepackage{threeparttable}
\usepackage[table]{xcolor}
\usepackage{mathtools}

\ifCLASSINFOpdf
 \usepackage[pdftex]{graphicx}
\else
\fi

\usepackage{amsmath}
\usepackage{xcolor}
\usepackage{algorithm}
\usepackage{algorithmicx}
\usepackage{amsmath,algorithm,tabularx}
\usepackage{algpseudocode}
\usepackage{amssymb}
\usepackage{multirow}
\usepackage{makecell}
\usepackage{caption}
\usepackage[absolute]{textpos}
\usepackage{array}
\newcolumntype{P}[1]{>{\centering\arraybackslash}p{#1}}
\newcolumntype{C}{>{\centering\arraybackslash} m{6cm} }
\usepackage{url}
\makeatletter
\newcommand{\multiline}[1]{%
	\begin{tabularx}{\dimexpr\linewidth-\ALG@thistlm}[t]{@{}X@{}}
		#1
	\end{tabularx}
}
\makeatother

\begin{document}
	\begin{textblock}{14}(1,0.1)
		\noindent S. Sun and H. Yan, ``Channel estimation for reconfigurable intelligent surface-assisted wireless communications considering Doppler effect," \textit{IEEE Wireless Communications Letters}, DOI: 10.1109/LWC.2020.3044004.
	\end{textblock}

\title{Channel Estimation for Reconfigurable Intelligent Surface-Assisted Wireless Communications Considering Doppler Effect}
\author{Shu~Sun and~Hangsong~Yan
\thanks{The authors are with the NYU WIRELESS research center and Tandon School of Engineering, New York University, Brooklyn, NY 11201 USA (e-mail: \{ss7152, hy942\}@nyu.edu).}}

\maketitle

\begin{abstract}
In wireless systems aided by reconfigurable intelligent surfaces (RISs), channel state information plays a pivotal role in achieving the performance gain of RISs. \textcolor{black}{Mobility renders accurate channel estimation (CE) more challenging due to the Doppler effect.} In this letter, we propose \textcolor{black}{two practical wideband CE schemes incorporating Doppler shift adjustment (DSA) for multi-path and single-path propagation environments, respectively, for RIS-assisted communication with passive reflecting elements. For the multi-path scenario, ordinary CE is first executed assuming quasi-static channels, followed by DSA realized via joint RIS reflection pattern selection and transformations between frequency and time domains. The proposed CE} necessitates only one more symbol incurring negligible extra overhead compared with the number of symbols required for the original CE. \textcolor{black}{For the single-path case which is especially applicable to millimeter-wave and terahertz systems, a novel low-complexity CE method is devised capitalizing on the form of the array factors at the RIS}. Simulation results demonstrate \textcolor{black}{that the proposed algorithms yield high CE accuracy and achievable rate with low complexity, and outperform representative benchmark schemes.}
\end{abstract}


\IEEEpeerreviewmaketitle

\section{Introduction}
\IEEEPARstart{R}{ECONFIGURABLE} intelligent surfaces (RISs), also known as holographic multiple-input-multiple-output (MIMO), intelligent beamforming metasurfaces, and intelligent reflecting surfaces (when used for reflection), among others, have recently stimulated an upsurge in research on their applications in wireless communications, due to their capability of smartly sensing and controlling the propagation environment~\cite{ElMossallamy20TCCN,Akyildiz18CM,Zhao20arXiv,Yu20JSAC}. An RIS comprises sub-wavelength metallic or dielectric scattering elements, which is capable of shaping electromagnetic waves via anomalous reflection, refraction, absorption, and/or polarization~\cite{ElMossallamy20TCCN,Akyildiz18CM,Sun_JOSA}. By appropriately and dynamically adjusting the amplitude and/or phase of each of the RIS elements based on the propagation environment, wireless signals can be coherently combined and steered to desired directions. Achieving the full potential of RISs requires the acquisition of accurate channel state information (CSI), which, however, usually incurs considerable overhead stemming from the large number of elements at an RIS. 

Pioneering works have been conducted to tackle the channel estimation (CE) issue in RIS-assisted systems. For instance, the authors in~\cite{Jung20TWC} have employed RISs where all or a portion of the elements are active to facilitate CE. In contrast to active RIS elements, purely passive RIS elements are more energy-efficient and cost-effective. CSI estimation algorithms for RISs with all passive elements are presented in~\cite{Wang20SPL,Wang20TWC}. Nevertheless, these works and analyzes are aimed for narrow-band wireless systems, while 5G and future 6G communications are likely to conduct broadband deployment. Wideband CE for RIS-aided systems is studied in~\cite{Zheng20WCL}, where the cascaded user-RIS-base-station channel is estimated by joint design of pilot sequences and the reflection pattern at the RIS. Furthermore, the authors in~\cite{Wan20ICC} leveraged compressive sensing to estimate the broadband channel for millimeter-wave (mmWave) systems. All the investigations above assume a quasi-static channel that remains approximately constant during CSI estimation, which, in reality, may not hold owing to motions at the communication ends and/or in the environment. \textcolor{black}{A compensation approach was proposed in~\cite{Matthiesen20WCL} to prevent RIS-induced additional frequency components. Furthermore, CE for RIS-assisted high-mobility communications was investigated in~\cite{Huang20arXiv} under the single-path assumption for the RIS reflecting link.}

In this letter, \textcolor{black}{we propose practical CE methods taking into account the Doppler effect in RIS-assisted wideband wireless systems for both multi-path and single-path scenarios. The main contributions of this letter are two-fold: First, we adopt the basic quasi-static multi-path CE framework proposed in~\cite{Zheng20WCL} and design a novel mechanism to adjust Doppler-induced distortions in estimated signals via frequency- and time-domain transformations. Second, we design a new low-complexity CE scheme leveraging the form of element array factors at the RIS for the single-path scenario which is typical in mmWave and terahertz (THz) communication systems. Numerical results show that the proposed schemes can effectively estimate the channel under UE mobility and outperform state-of-the-art benchmark strategies.} 

\vspace{-0.5cm}
\section{System Model and Problem Formulation}
\subsection{System Model}
\begin{figure}
	\centering
	\includegraphics[width=0.7\columnwidth]{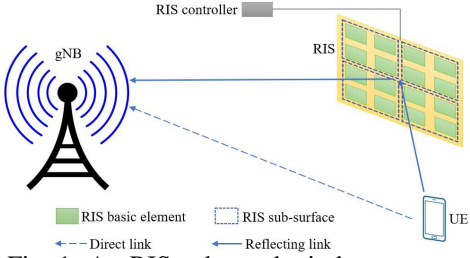}
	\caption{An RIS-enhanced wireless system.}
	\label{fig:inputOutput}	
\end{figure}

We consider an uplink point-to-point communication system assisted by an RIS, which consists of an $N_\text{G}$-antenna next-generation nodeB (gNB), an $N_\text{U}$-antenna user equipment (UE), and a two-dimensional passive RIS comprising $N_\text{R}$ reconfigurable elements, as illustrated in Fig.~\ref{fig:inputOutput}, where the RIS is employed to enhance the communication by creating a reflecting link between the gNB and the UE. To reduce the complexity of CE and hardware cost, and given the high correlation of close-by elements in the RIS due to sub-wavelength spacing, the entire RIS is divided into $M$ sub-surfaces, each of which contains $N_\text{R}/M$ (set to be an integer) adjacent elements. 

In this work, frequency-selective fading channels are considered to represent many practical wireless channels. The total bandwidth allocated to the UE is divided into $N$ sub-carriers, or equivalently, $N_\text{B}=N/12$ resource blocks (RBs). Without loss of generality, we assume that the maximum delay spread is of $L$ taps in the time domain for the baseband equivalent channels of both the UE-gNB direct link and the UE-RIS-gNB reflecting link. For the link between the $u$-th $(\forall u=0,..,N_\text{U}-1)$ antenna at the UE and the $g$-th $(\forall g=0,..,N_\text{G}-1)$ antenna at the gNB, each transmitted orthogonal frequency-division multiplexing (OFDM) symbol is denoted as $\textbf{x}=[x_0,...,x_{N-1}]^T\in\mathbb{C}^{N\times1}$ (where the subscripts for antenna indices are omitted for simplicity), which is first transformed into the time domain through an $N$-point inverse discrete Fourier transform (IDFT), followed by the appending of a cyclic prefix (CP) of length $L_\text{CP}$ that is assumed to be larger than $L$. At the gNB, after removing the CP and performing the $N$-point discrete Fourier transform (DFT), the baseband received signal in the frequency domain is given by (the symbol index is omitted herein)
\begin{equation}\label{eq:y1}
\textbf{y}=\Bigg(\bigg(\sum_{m=0}^{M-1}\textbf{h}_{\text{RG},m}\circ\phi_m\textbf{h}_{\text{UR},m}\bigg)+\textbf{h}_\text{UG}\Bigg)\circ\textbf{x}+\textbf{w}
\end{equation}

\noindent where $\circ$ denotes the Hadamard product, $\textbf{y}\in\mathbb{C}^{N\times1}$ is the received OFDM symbol, $\textbf{h}_{\text{RG},m}\in\mathbb{C}^{N\times1}$ is the channel frequency response (CFR) of the RIS-gNB link associated with the $m$-th sub-surface at the RIS, $\phi_m=\alpha_me^{j\varphi_m}$ is the reflection coefficient of the $m$-th sub-surface, $\textbf{h}_{\text{UR},m}\in\mathbb{C}^{N\times1}$ is the CFR of the UE-RIS link for the $m$-th sub-surface, $\textbf{h}_\text{UG}\in\mathbb{C}^{N\times1}$ represents the CFR of the UE-gNB direct link, and $\textbf{w}\in\mathbb{C}^{N\times1}$ stands for the received additive white Gaussian noise (AWGN) with $\textbf{w}\thicksim\mathcal{CN}(\textbf{0},\sigma^2\textbf{I}_N)$. To maximize the reflection power of the RIS and simplify its hardware design, we set $\alpha_m=1,\forall m=0,...,M-1$.

Let $\textbf{h}_{\text{URG},m}=\textbf{h}_{\text{RG},m}\circ\textbf{h}_{\text{UR},m}\in\mathbb{C}^{N\times1}$ denote the effective cascaded CFR of the reflecting link for the $m$-th sub-surface (without the effect of RIS phase shift), and $\textbf{X}=\text{diag}(\textbf{x})\in\mathbb{C}^{N\times N}$ the diagonal matrix of $\textbf{x}$, then (\ref{eq:y1}) can be recast as 
\begin{equation}\label{eq:y2}
\textbf{y}=\textbf{X}\Bigg(\bigg(\sum_{m=0}^{M-1}\phi_m\textbf{h}_{\text{URG},m}\bigg)+\textbf{h}_\text{UG}\Bigg)+\textbf{w}
\end{equation}

\noindent By concatenating $\textbf{h}_{\text{URG},m}$ into $\textbf{H}_\text{URG}=[\textbf{h}_{\text{URG},0},...,\textbf{h}_{\text{URG},M-1}]\in\mathbb{C}^{N\times M}$ and set $\mathbf{\Phi}=[\phi_0,...,\phi_{M-1}]^T\in\mathbb{C}^{M\times1}$, (\ref{eq:y2}) can be expressed more succinctly as
\begin{equation}\label{eq:y3}
\mathbf{y}=\textbf{X}(\underbrace{\textbf{H}_\text{URG}\mathbf{\Phi}+\textbf{h}_\text{UG}}_\textbf{h})+\textbf{w}=\textbf{X}\textbf{h}+\textbf{w}
\end{equation}
\vspace{-0.7cm}
\subsection{Wideband Channel Estimation}
With the simple form in (\ref{eq:y3}), $\textbf{h}$ can be estimated via various conventional CE techniques. One popular and standard approach is to employ pilot signals to estimate the uplink channel to facilitate the subsequent resource allocation and scheduling. Denoting the estimation of $\textbf{h}$ as $\hat{\textbf{h}}$, $\textbf{H}_\text{URG}$ and $\textbf{h}_\text{UG}$ need to be resolved after $\hat{\textbf{h}}$ is obtained, which can be realized by proper design of the reflection pattern $\mathbf{\Theta}$ at the RIS,\footnote{\textcolor{black}{The reflection pattern has to remain constant over sub-carriers since only one pattern can be configured each time, although the optimal reflection pattern for a wideband channel may vary over sub-carriers. In the beamforming strategy employed from Section III-B of~\cite{Zheng20WCL}, the reflection coefficients are optimized to maximize the strongest time-domain path gain.}} such as using the method in~\cite{Zheng20WCL}. With a pre-designed $\mathbf{\Theta}$, the estimation of $\textbf{H}_\text{URG}$ and $\textbf{h}_\text{UG}$ can be obtained as follows
\begin{equation}\label{eq:HHat1}
[\hat{\textbf{h}}_\text{UG},\hat{\textbf{H}}_\text{URG}]=\hat{\textbf{H}}\mathbf{\Theta}^{-1}
\end{equation}

\noindent where $\hat{\textbf{H}}=[\hat{\textbf{h}}^{(0)},...,\hat{\textbf{h}}^{(M)}]\in\mathbb{C}^{N\times(M+1)}$ with $\hat{\textbf{h}}^{(i)}$ representing the estimation of the superimposed CFR in the $i$-th symbol. The proposed CE scheme in~\cite{Zheng20WCL} outperforms some prior methods in the literature. A crucial assumption in~\cite{Zheng20WCL} is that the channel is quasi-static within the transmission frame such that $\textbf{H}_\text{URG}$ and $\textbf{h}_\text{UG}$ remain approximately constant. However, since the number of symbols required for CE is $M+1$, the estimation duration is long for large values of  $M$ and the assumption of quasi-static channel may not hold in practical scenarios, hence degrading CE performance and the resultant achievable rate.  

\vspace{-0.4cm}
\subsection{Problem Formulation}
\vspace{-0.1cm}
Consider the situation where the UE is moving at a certain velocity $v$, the induced phase change for a path is $\Delta\psi=2\pi v\Delta t\text{cos}(\theta)/\lambda$~\cite{Rappaport02Book} , with $\Delta t$ denoting the time passed with respect to a reference moment, $\lambda$ the wavelength, and $\theta$ the traveling angle of a path with respect to the moving direction. For a single path, the Doppler effect gives rise to a phase change proportional to $\Delta t$; while for a wideband signal containing more than one path, it may encounter both phase and amplitude alterations due to the superposition of the phase-shifted paths. If the UE is moving, the true CFR $\textbf{H}_\text{URG}$ and $\textbf{h}_\text{UG}$ will vary over the $M+1$ symbols during CE due to Doppler shift. Based on (\ref{eq:y3}), the true CFR in the $i$-th symbol can be rewritten as 
\begin{equation}\label{eq:hi1}
\textbf{h}^{(i)}=\textbf{H}^{(i)}_{\text{URG}}\mathbf{\Phi}^{(i)}+\textbf{h}^{(i)}_{\text{UG}}
\end{equation}

\noindent Without loss of generality, we take the 1st symbol as the reference symbol, then the Doppler effect on the subsequent symbols can be modeled as 
\begin{equation}\label{eq:hi2}
\begin{split}
\textbf{h}^{(i)}&=\Big(\textbf{H}^{(1)}_{\text{URG}}+\mathbf{\Xi}^{(i-1)}\Big)\mathbf{\Phi}^{(i)}+\Big(\textbf{h}^{(1)}_{\text{UG}}+\boldsymbol{\epsilon}^{(i-1)}\Big) \\
&=\overrightarrow{\textbf{H}}^{(1)}_{\text{UG}}\overrightarrow{\mathbf{\Phi}}^{(i)}+\mathbf{\Upsilon}^{(i-1)}\overrightarrow{\mathbf{\Phi}}^{(i)}
\end{split}
\end{equation}

\noindent where $\mathbf{\Xi}^{(i-1)}$ and $\boldsymbol{\epsilon}^{(i-1)}$ represent the variation of $\textbf{H}^{(1)}_{\text{URG}}$ and $\textbf{h}^{(1)}_{\text{UG}}$ in the $i$-th symbol with respect to the 1st symbol, respectively, $\overrightarrow{\textbf{H}}^{(1)}_{\text{UG}}=[\textbf{h}^{(1)}_{\text{UG}},\textbf{H}^{(1)}_{\text{URG}}]$, $\mathbf{\Upsilon}^{(i-1)}=[\boldsymbol{\epsilon}^{(i-1)},\mathbf{\Xi}^{(i-1)}]$, and $\overrightarrow{\mathbf{\Phi}}^{(i)}=\left[\begin{array}{c}1\\\mathbf{\Phi}^{(i)}\end{array}\right]$. Stacking $\textbf{h}^{(i)}$ in (\ref{eq:hi2}) over $M+1$ consecutive symbols yields  
\begin{equation}\label{eq:H1}
\textbf{H}=\overrightarrow{\textbf{H}}^{(1)}_{\text{UG}}\mathbf{\Theta}+\left[\mathbf{\Upsilon}^{(0)}\overrightarrow{\mathbf{\Phi}}^{(1)},...,\mathbf{\Upsilon}^{(M)}\overrightarrow{\mathbf{\Phi}}^{(M+1)}\right]
\end{equation}

\noindent Therefore
\begin{equation}\label{eq:HHat2}
\hat{\textbf{H}}\mathbf{\Theta}^{-1}=\overrightarrow{\textbf{H}}^{(1)}_{\text{UG}}+\left[\mathbf{\Upsilon}^{(0)}\overrightarrow{\mathbf{\Phi}}^{(1)},...,\mathbf{\Upsilon}^{(M)}\overrightarrow{\mathbf{\Phi}}^{(M+1)}\right]\mathbf{\Theta}^{-1}+\textbf{V}\mathbf{\Theta}^{-1}
\end{equation}

\noindent where $\mathbf{\Theta}$ is the pre-designed reflection pattern used in (\ref{eq:HHat1}), and $\textbf{V}\in\mathbb{C}^{N\times(M+1)}$ is the noise term. It is evident from (\ref{eq:HHat2}) that with Doppler effect, the equivalent CFR estimated by (\ref{eq:HHat1}) is $\overrightarrow{\textbf{H}}^{(1)}_{\text{UG}}+\left[\mathbf{\Upsilon}^{(0)}\overrightarrow{\mathbf{\Phi}}^{(1)},...,\mathbf{\Upsilon}^{(M)}\overrightarrow{\mathbf{\Phi}}^{(M+1)}\right]\mathbf{\Theta}^{-1}$ instead of $\overrightarrow{\textbf{H}}^{(1)}_{\text{UG}}$, resulting in non-negligible estimation error. The problem now is how to account for the Doppler effect in (\ref{eq:HHat2}) to obtain more accurate estimation of $\overrightarrow{\textbf{H}}^{(i)}_{\text{UG}}$, $\forall i=1,...,M+1$.

\section{Proposed \textcolor{black}{CE for Multi-Path Scenario}}\label{sec:ProDSA}
In this section, \textcolor{black}{we present a CE mechanism with Doppler shift adjustment (DSA) to address the issue mentioned above for the multi-path scenario}. As shown in (\ref{eq:HHat2}), the estimation error incurred by the Doppler effect is tangled with the pre-designed reflection pattern $\mathbf{\Theta}$ which needs to be separated out for the evaluation of Doppler shift alone. To this end, it is necessary to add one more symbol that shares the reflection pattern with one of the $M+1$ symbols adopted for CE, and the extra symbol should be adjacent to the pattern-sharing symbol to precisely capture the CFR variation over one symbol. For ease of exposition, the extra symbol is set as the 0th symbol and the associated reflection pattern is $\mathbf{\Phi}^{(0)}=\mathbf{\Phi}^{(1)}=[1,...,1]^T\in\mathbb{C}^{M\times1}$, i.e., it shares the reflection pattern with the 1st symbol.  

Inspired by the fact that the motion of the UE leads to a phase shift for each path in the channel impulse response (CIR), it is easier to adjust the Doppler effect in the time domain as opposed to the frequency domain. Thus, for the estimated superimposed CFR in the $i$-th symbol $\hat{\textbf{h}}^{(i)}$ ($i\in\left\{0,...,M+1\right\}$), an $N$-point IDFT is performed to obtain its CIR $\hat{\textbf{g}}^{(i)}\in\mathbb{C}^{N\times1}$ given by (with a slight abuse of notations)
\begin{equation}\label{eq:gi1}
[\hat{\textbf{g}}^{(i)}]_p=\sum_{k=0}^{N-1}a_k^{(i)}e^{j\beta_k^{(i)}}\delta(p-k), p=0,1,...,N-1
\end{equation}

\noindent where $p$ is the path index, $a_k^{(i)}$ and $\beta_k^{(i)}$ denote the amplitude and phase of the $k$-th path in $\hat{\textbf{g}}^{(i)}$, respectively. The position and value of $a_k^{(i)}$ are assumed to be consistent over the $M+2$ symbols, which is reasonable since the Doppler effect only alters the phase of each path. To lower the computational complexity and mitigate the impact of random noise, only the paths whose amplitudes are no smaller than a pre-defined threshold $\varpi$ (e.g., 0.1) with respect to the maximum amplitude are selected in $\hat{\textbf{g}}^{(0)}$ for succeeding phase adjustment, and the set of such paths is denoted as $\mathcal{P}$. Then (\ref{eq:gi1}) can be re-organized as 
\vspace{-5pt}
\begin{equation}\label{eq:gi2}
[\hat{\textbf{g}}^{(i)}]_p=\sum_{k\in\mathcal{P}}a_k^{(i)}e^{j\beta_k^{(i)}}\delta(p-k)+\sum_{k^\prime\notin\mathcal{P}}a_{k^\prime}^{(i)}e^{j\beta_{k^\prime}^{(i)}}\delta(p-k^\prime)
\end{equation}

\noindent The phase of the $k$-th ($\forall k\in\mathcal{P}$) path is compared between $\hat{\textbf{g}}^{(1)}$ and $\hat{\textbf{g}}^{(0)}$, i.e., $\Delta\beta_k=\beta_k^{(1)}-\beta_k^{(0)}$, which is regarded as the phase difference between adjacent symbols of the $k$-th path. Next, to estimate $\overrightarrow{\textbf{H}}^{(q)}_{\text{UG}}$ ($q\in\left\{1,...,M+1\right\}$) \textcolor{black}{using all the $\hat{\textbf{g}}^{(i)}$ ($\forall i=1,...,M+1$)}, the phase of the $k$-th ($\forall k\in\mathcal{P}$) path in each $\hat{\textbf{g}}^{(i)}$ is adjusted based on $\Delta\beta_k$
\begin{equation}\label{eq:gi3}
\begin{split}
[\hat{\textbf{g}}^{(i)}]_p=&\sum_{k\in\mathcal{P}}a_k^{(i)}e^{j\left(\beta_k^{(i)}-(i-q)\Delta\beta_k\right)}\delta(p-k)\\
&+\sum_{k^\prime\notin\mathcal{P}}a_{k^\prime}^{(i)}e^{j\beta_{k^\prime}^{(i)}}\delta(p-k^\prime)
\end{split}
\end{equation}

\noindent \textcolor{black}{where the term $-(i-q)\Delta\beta_k$ accounts for the phase shift caused by the Doppler effect on the $k$-th path between the $i$-th symbol and the $q$-th symbol.} This way, $\hat{\textbf{g}}^{(i)}$ is adjusted as if it were the CIR (including the reflection pattern at the RIS) associated with the $i$-th symbol without Doppler shift \textcolor{black}{between the $i$-th symbol and the $q$-th symbol}. $\hat{\textbf{g}}^{(i)}$ is then converted back to the frequency domain via the $N$-point DFT to yield $\hat{\textbf{h}}^{(i)}_{q,\text{adj}}$. Finally, the adjusted estimation of $\textbf{h}^{(q)}_{\text{UG}}$ and $\textbf{H}^{(q)}_{\text{URG}}$ is given by 
\begin{equation}\label{eq:HHat3}
\left[\hat{\textbf{h}}^{(q)}_{\text{UG}},\hat{\textbf{H}}^{(q)}_\text{URG}\right]=\hat{\textbf{H}}_{q,\text{adj}}\mathbf{\Theta}^{-1}
\end{equation}

\noindent where $\hat{\textbf{H}}_{q,\text{adj}}=\left[\hat{\textbf{h}}^{(1)}_{q,\text{adj}},...,\hat{\textbf{h}}^{(M+1)}_{q,\text{adj}}\right]\in\mathbb{C}^{N\times(M+1)}$. Essentially, more accurate estimation of $\textbf{h}^{(q)}_{\text{UG}}$ and $\textbf{H}^{(q)}_{\text{URG}}~(\forall q=1,...,M+1)$ is realized by reconstructing $\hat{\textbf{h}}^{(i)}_{q,\text{adj}}$ (and hence $\hat{\textbf{H}}_{q,\text{adj}}$) via phase adjustment in the corresponding CIRs.\footnote{\textcolor{black}{The proposed DSA is applicable to any reflection pattern $\mathbf{\Theta}$ as long as $\text{rank}(\mathbf{\Theta})=M+1$ and each reflection coefficient has unit modulus. Nevertheless, to avoid the high-complexity inversion operation and to minimize the mean square estimation error of $\left[\hat{\textbf{h}}^{(q)}_{\text{UG}},\hat{\textbf{H}}^{(q)}_\text{URG}\right]$, we adopt the reflection pattern in~\cite{Zheng20WCL} where $\mathbf{\Theta}$ is an $(M+1)\times(M+1)$ DFT matrix with $[\mathbf{\Theta}]_{p.q}=e^{-j\frac{2\pi pq}{M+1}},~0\leq p,q\leq M$. More details can be found in Section III of~\cite{Zheng20WCL}.}}

\vspace{- 5pt}
\section{\textcolor{black}{Proposed CE} for Single-Path Scenario}
In this section, we propose a novel and low-complexity CE scheme when the channel has (approximately) solely the line-of-sight (LoS) path in the UE-RIS and RIS-gNB links. This scenario is particularly applicable to mmWave and THz communication channels due to severe propagation loss caused by atmospheric attenuation, as well as reflection, scattering, and penetration loss at those high carrier frequencies~\cite{Akyildiz18CM}. For ease of exposition, the UE-gNB direct link is neglected for the subsequent analysis, yet the extension to the scenario with the direct link is straightforward. Under this condition, the formulation in (\ref{eq:y3}) for each sub-carrier can be simplified to  
\begin{equation}\label{eq:y}
y=x\sum_{m=0}^{M-1}\phi_mh_{\text{URG},m}+w=x\sum_{m=0}^{M-1}[\tilde{\textbf{h}}_\text{RG}^*\circ\tilde{\textbf{h}}_\text{UR}]_m+w
\end{equation}

\noindent where we set $\phi_m=1~(\forall m=0,...,M-1)$, $\tilde{\textbf{h}}_\text{UR}\in\mathbb{C}^{M\times1}$ and $\tilde{\textbf{h}}_\text{RG}\in\mathbb{C}^{M\times1}$ are the CFRs for the UE-RIS and RIS-gNB links, respectively, and $\tilde{\textbf{h}}_\text{RG}^*$ denotes the conjugate of $\tilde{\textbf{h}}_\text{RG}$. The target herein is to estimate $\tilde{\textbf{h}}_\text{RG}^*\circ\tilde{\textbf{h}}_\text{UR}$ for all sub-surfaces at the RIS. For each sub-carrier and each pair of the UE and gNB antennas, the mmWave propagation environment is well characterized by the Saleh-Valenzuela model depicted as~\cite{Wang20SPL} (with a slight abuse of notations)  
\vspace{-5pt}
\begin{equation}\label{eq:hUR1}
\tilde{\textbf{h}}_\text{UR}=\sqrt{\frac{M}{P}}\sum_{p=0}^{P-1}\alpha_p\textbf{a}_\text{R}(\theta_p,\varphi_p),\tilde{\textbf{h}}_\text{RG}=\sqrt{\frac{M}{Q}}\sum_{q=0}^{Q-1}\varrho_q\textbf{a}_\text{R}(\vartheta_q,\psi_q)
\end{equation}

\noindent where $\alpha_p$ and $\theta_p (\varphi_p)$ denote the complex gain and azimuth (elevation) angle of arrival (AoA) of the $p$-th path of the UE-RIS channel, respectively. Likewise, $\varrho_q$ and $\vartheta_q (\psi_q)$ represent the complex gain and azimuth (elevation) angle of departure (AoD) of the $q$-th path of the RIS-gNB channel. Additionally, $\textbf{a}_\text{R}$ denotes the array response vector at the RIS. For a uniform square array with $\sqrt{M}\times\sqrt{M}$ elements, $\textbf{a}_\text{R}$ is given by
\begin{equation}\label{eq:aR}
\begin{split}
\textbf{a}_\text{R}(\theta,\varphi)=&\frac{1}{\sqrt{M}}\Big[1,...,e^{j\frac{2\pi}{\lambda}d(u\text{sin}\theta\text{sin}\varphi+\tilde{u}\text{cos}\varphi)},...,\\
&e^{j\frac{2\pi}{\lambda}d((\sqrt{M}-1)\text{sin}\theta\text{sin}\varphi+(\sqrt{M}-1)\text{cos}\varphi)}\Big]^T
\end{split}
\end{equation}

\vspace{-7pt}
\noindent where $d$ is the antenna spacing, and $0\leq u,\tilde{u}<\sqrt{M}$ are the element indices in the 2D plane. When there is one dominant path (denoted as the 0-th path without loss of generality) whose amplitude is substantially larger than all the other paths, (\ref{eq:hUR1}) can be approximated as 
\vspace{-6.6pt}
\begin{equation}\label{eq:hUR2}
\tilde{\textbf{h}}_\text{UR}\approx\sqrt{\frac{M}{P}}\alpha_0\textbf{a}_\text{R}(\theta_0,\varphi_0),\tilde{\textbf{h}}_\text{RG}\approx\sqrt{\frac{M}{Q}}\varrho_0\textbf{a}_\text{R}(\vartheta_0,\psi_0)
\end{equation}

\noindent such that 
\begin{equation}\label{eq:hRGhUR}
\tilde{\textbf{h}}_\text{RG}^*\circ\tilde{\textbf{h}}_\text{UR}\approx A[1,...,a^ub^{\tilde{u}},...,a^{\sqrt{M}-1}b^{\sqrt{M}-1}]^T
\end{equation}

\noindent where $A=\alpha_0\varrho_0/\sqrt{PQ}$, $a=e^{j\frac{2\pi}{\lambda}d(\text{sin}\theta_0\text{sin}\varphi_0-\text{sin}\vartheta_0\text{sin}\psi_0)}$, and $b=e^{j\frac{2\pi}{\lambda}d(\text{cos}\varphi_0-\text{cos}\psi_0)}$ are constants relying merely on the complex gains, and azimuth and elevation AoA and AoD of the dominant path. (\ref{eq:hRGhUR}) indicates that $\tilde{\textbf{h}}_\text{RG}^*\circ\tilde{\textbf{h}}_\text{UR}$ can be calculated if $A$, $a$, and $b$ are known. Theoretically only three pilot symbols are needed to acquire $A$, $a$, and $b$, while DSA necessitates one more symbol. Therefore, only four pilot symbols are needed to obtain $\tilde{\textbf{h}}_\text{RG}^*\circ\tilde{\textbf{h}}_\text{UR}$ regardless of the value of $M$, which significantly reduces the CE time and hence the overall latency as compared with the situation in Section~\ref{sec:ProDSA}.   

The proposed CE approach is detailed below. When sending the first two pilot symbols, all the sub-surfaces at the RIS are turned off except the 0-th one ($u=\tilde{u}=0$ in (\ref{eq:hRGhUR})), which gives $\hat{A}^{(0)}=y^{(0)}/x,~\hat{A}^{(1)}=y^{(1)}/x$. Since there is only one path, the Doppler phase shift between adjacent symbols can be conveniently obtained by comparing the phases of $\hat{A}^{(1)}$ and $\hat{A}^{(0)}$, i.e.,
\begin{equation}\label{eq:deltaZeta}
\Delta\zeta=\angle\hat{A}^{(1)}-\angle\hat{A}^{(0)}
\end{equation}

\noindent For the next two pilot symbols, all the sub-surfaces at the RIS are turned off except the one corresponding to $\langle u=1,\tilde{u}=0\rangle$ and $\langle u=0,\tilde{u}=1\rangle$ in (\ref{eq:hRGhUR}), respectively, yielding 
\begin{equation}\label{eq:Ahat1}
\hat{A}^{(2)}\hat{a}=\hat{A}^{(1)}e^{j\Delta\zeta}\hat{a}=\frac{y^{(2)}}{x},~\hat{A}^{(3)}\hat{b}=\hat{A}^{(1)}e^{j2\Delta\zeta}\hat{b}=\frac{y^{(3)}}{x}
\end{equation}

\noindent where (\ref{eq:deltaZeta}) has been invoked. Accordingly, 
\begin{equation}\label{eq:Ahat2}
\hat{A}^{(i)}=\frac{e^{j(i-1)\Delta\zeta}y^{(1)}}{x},~\hat{a}=\frac{e^{-j\Delta\zeta}y^{(2)}}{y^{(1)}},~\hat{b}=\frac{e^{-j2\Delta\zeta}y^{(3)}}{y^{(1)}}
\end{equation}

\noindent Consequently, all the elements in $\tilde{\textbf{h}}_\text{RG}^*\circ\tilde{\textbf{h}}_\text{UR}$ can be derived based on (\ref{eq:deltaZeta}) and (\ref{eq:Ahat2}), hence solving the CE problem.   

\section{Numerical Results}
In this section, we provide simulation results for performance evaluation of the proposed multi-path and single-path CE methods. A uniform square array is equipped at the RIS consisting of 576 elements with $\lambda/4$ spacing, and one antenna is assumed at the gNB and UE. The center carrier frequency is 28 GHz, the number of RBs is 16 with 30 kHz sub-carrier spacing, the noise power is about -106 dBm, and 40 dB processing gain is presumed at the gNB which is practical for mmWave receivers to enhance the link budget~\cite{Sun17SmallScale}. The Zadoff-Chu sequence is utilized as the pilot sequence. For both the direct and reflecting links in the multi-path CE, 6-tap channels are assumed with $\eta$ being the ratio of the total power of non-dominant paths to that of the dominant path. The UE-RIS and RIS-gNB distances are 5 m and 50 m, respectively, while the UE location is randomly selected in each simulation run. The path loss exponents for the UE-RIS, RIS-gNB, and UE-gNB links are set to 2.0, 2.1, and 3.5, respectively. \textcolor{black}{The UE velocity is set to 10 m/s. The achievable rate is obtained using the beamforming approach in~\cite{Zheng20WCL}. The proposed algorithms are compared against two multi-path benchmark schemes and one single-path baseline method:}
\begin{enumerate}[label={\arabic*)}]
\item \textcolor{black}{Multi-path benchmark 1: The conventional cascaded CE scheme in~\cite{Zheng20WCL} without considering the Doppler effect.}

\item \textcolor{black}{Multi-path benchmark 2: A simplified DSA strategy where the phase is adjusted based on only the strongest time-domain path, which can be deemed a special case of our proposed multi-path scheme with the amplitude threshold $\varpi$ set to 1.}

\item \textcolor{black}{Single-path benchmark: The LoS single-path CE algorithm recently proposed in~\cite{Huang20arXiv}.}
\end{enumerate}
\textcolor{black}{The computational complexity and minimum CE latency of all the aforementioned schemes are analyzed in Table~\ref{tbl:complexity}. For the schemes with DSA, the proposed ones have comparable or even lower computational complexity and CE latency than the corresponding baselines. Moreover, for the multi-path scenario, both baseline 2 and the proposed approach necessitate only one more symbol compared with baseline 1 without DSA.}

\begin{table}[!t]\color{black}
	\centering
	\begin{threeparttable}
	\renewcommand{\arraystretch}{1.1}
	\caption{\textcolor{black}{Complexity and CE Latency Comparison}\tnote{*}}
	\label{tbl:complexity}
	\begin{tabular}{|m{1.1cm}||m{1cm}|m{1.25cm}|m{1.25cm}|m{0.7cm}|m{0.85cm}|}
		\hline
		\multirow{2}{*}{} & \multicolumn{3}{c|}{Multi-Path} & \multicolumn{2}{c|}{Single-Path} \\
		\cline{2-6}
		& \makecell{BM1~\cite{Zheng20WCL}\\(No DSA)} & BM2 & Proposed & \makecell{BM\\\cite{Huang20arXiv}} & Proposed\\
		\hline 
		\makecell{Complexity} & \makecell{$\mathcal{O}(M^2N)$}& \makecell{$\mathcal{O}(M^2N+$\\$MN\log N)$} & \makecell{$\mathcal{O}(M^2N+$\\$MN\log N)$} & $\mathcal{O}(M^3)$ & $\mathcal{O}(M)$\\
		\hline
		\makecell{Minimum \\CE latency} & $M+1$ & $M+2$ & $M+2$ & $\tau$ & 4\\
		\hline
	\end{tabular}
\begin{tablenotes}\footnotesize
	\item[*] \textcolor{black}{BM is short for benchmark; Minimum CE latency equals the number of symbols needed for CE; $\tau\in\mathbb{Z}^+$ denotes the number of pilot symbols. For multi-path CE, the proposed scheme actually has slightly higher complexity than BM2, but the extra complexity is negligible compared with $\mathcal{O}(M^2N+MN\log N)$.}
\end{tablenotes}
\end{threeparttable}
\end{table}

\begin{figure}
	\centering
	\includegraphics[width=\columnwidth]{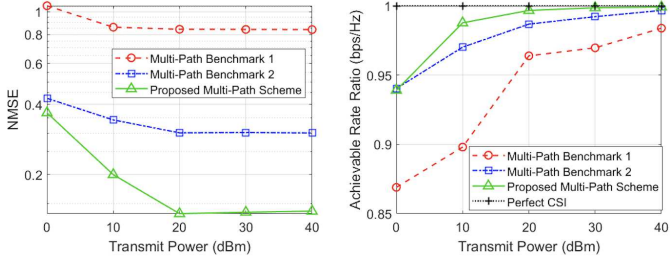}
	\caption{\textcolor{black}{Multi-path NMSE and achievable rate ratio versus transmit power with $M=16$ and $\eta=0.1$.}}
	\label{fig:seRatioVsPt_v10_pRatio0_1_M16}	
\end{figure}

\textcolor{black}{For the multi-path scenario, Fig.~\ref{fig:seRatioVsPt_v10_pRatio0_1_M16} illustrates the CE normalized mean square error (NMSE) (normalized to the channel gain) for the reflecting link and the achievable rate ratio against perfect CSI with $M=16$ and $\eta=0.1$. It can be observed that the proposed scheme always produces more accurate CE and higher achievable rate compared with both of the baseline counterparts. Specifically, the maximum achievable rate reduction against perfect CSI reaches 13\% in the first baseline method, while it is only 6\% using the proposed and the second baseline approaches. Moreover, take the 10 dBm transmit power for example, the proposed method renders about 10\% and 2\% achievable rate gain against the first and second baseline schemes, respectively, demonstrating the superior CE performance of the proposed scheme.}
\begin{figure}
	\centering
	\includegraphics[width=\columnwidth]{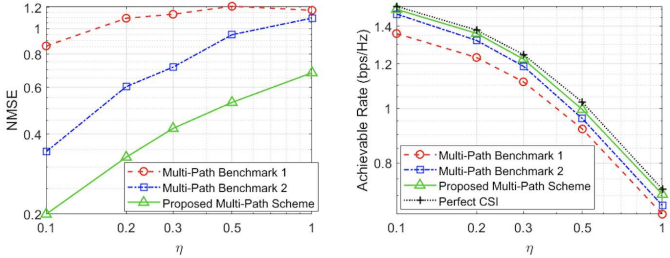}
	\caption{\textcolor{black}{Multi-path NMSE and achievable rate versus $\eta$ with $M=16$ and 10 dBm transmit power.}}
	\label{fig:r1_multipath_mnse_se_pRatio_M16_pt10}	
\end{figure}

\textcolor{black}{The effect of $\eta$, the ratio of the total power of non-dominant paths to that of the dominant one, is examined in Fig.~\ref{fig:r1_multipath_mnse_se_pRatio_M16_pt10}, which shows that the proposed scheme outperforms both of the baseline algorithms for a wide range of $\eta$ varying from 0.1 to 1, corroborating its robustness against $\eta$. Note that the achievable rate for all the schemes (including the perfect CSI case) decreases with $\eta$, due to the strongest-CIR-alignment reflection optimization scheme adopted from~\cite{Zheng20WCL}, such that the mismatch between the RIS phase pattern and the phases of all the paths increases with the power of non-dominant paths since the RIS phase is aligned to only the dominant path.}

\begin{figure}
	\centering
	\includegraphics[width=\columnwidth]{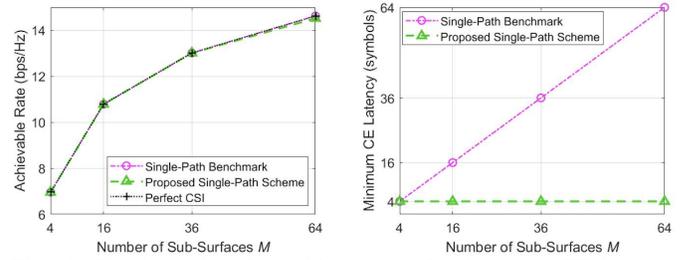}
	\caption{\textcolor{black}{Single-path achievable rate and minimum CE latency versus $M$ with 25 dBm transmit power.}}
	\label{fig:r1_mnse_se_M}	
\end{figure}

\textcolor{black}{Next, we investigate the single-path scenario, where the achievable rate and minimum CE latency are displayed in Fig.~\ref{fig:r1_mnse_se_M}. Note that theoretically the number of pilot symbols $\tau$ in the benchmark scheme can be flexible, but actually $\tau\geq M$ needs to hold given the pseudo inversion operation in (21) of~\cite{Huang20arXiv}, thus we set $\tau$ to $M$ herein. It is evident from Fig.~\ref{fig:r1_mnse_se_M} that the proposed scheme yields the same or only slightly lower achievable rate\footnote{\textcolor{black}{The slightly lower achievable rate is attributed to the fact that larger $M$ results in higher phase resolution, hence making CE more vulnerable to phase variation caused by noise.}} than the benchmark with $\tau=M$, while enjoying lower latency and complexity (see also Table~\ref{tbl:complexity}).}

\section{Conclusion}
In this letter, we have proposed \textcolor{black}{two practical CE schemes with DSA for both multi-path and single-path scenarios} in RIS-assisted wideband wireless systems. Simulation results \textcolor{black}{show that the proposed schemes are superior to representative baseline ones considering CE accuracy along with achievable rate, computational complexity, and CE latency.}

\ifCLASSOPTIONcaptionsoff
  \newpage
\fi
\bibliographystyle{IEEEtran}
\bibliography{Sun_WCL2020-1759}

\begin{thebibliography}{10}
\providecommand{\url}[1]{#1}
\csname url@samestyle\endcsname
\providecommand{\newblock}{\relax}
\providecommand{\bibinfo}[2]{#2}
\providecommand{\BIBentrySTDinterwordspacing}{\spaceskip=0pt\relax}
\providecommand{\BIBentryALTinterwordstretchfactor}{4}
\providecommand{\BIBentryALTinterwordspacing}{\spaceskip=\fontdimen2\font plus
\BIBentryALTinterwordstretchfactor\fontdimen3\font minus
  \fontdimen4\font\relax}
\providecommand{\BIBforeignlanguage}[2]{{%
\expandafter\ifx\csname l@#1\endcsname\relax
\typeout{** WARNING: IEEEtran.bst: No hyphenation pattern has been}%
\typeout{** loaded for the language `#1'. Using the pattern for}%
\typeout{** the default language instead.}%
\else
\language=\csname l@#1\endcsname
\fi
#2}}
\providecommand{\BIBdecl}{\relax}
\BIBdecl

\bibitem{ElMossallamy20TCCN}
M.~A. {ElMossallamy} \emph{et~al.}, ``Reconfigurable intelligent surfaces for
  wireless communications: Principles, challenges, and opportunities,''
  \emph{IEEE Transactions on Cognitive Communications and Networking}, vol.~6,
  no.~3, pp. 990--1002, Sep. 2020.

\bibitem{Akyildiz18CM}
I.~F. {Akyildiz} \emph{et~al.}, ``Combating the distance problem in the
  millimeter wave and terahertz frequency bands,'' \emph{IEEE Communications
  Magazine}, vol.~56, no.~6, pp. 102--108, Jun. 2018.

\bibitem{Zhao20arXiv}
\BIBentryALTinterwordspacing
M.~{Zhao} \emph{et~al.}, ``Two-timescale beamforming optimization for
  intelligent reflecting surface aided multiuser communication with {QoS}
  constraints.'' [Online]. Available: \url{https://arxiv.org/abs/2011.02237.}
\BIBentrySTDinterwordspacing

\bibitem{Yu20JSAC}
X.~{Yu} \emph{et~al.}, ``Robust and secure wireless communications via
  intelligent reflecting surfaces,'' \emph{IEEE Journal on Selected Areas in
  Communications}, vol.~38, no.~11, pp. 2637--2652, Nov. 2020.

\bibitem{Sun_JOSA}
S.~{Sun} \emph{et~al.}, ``Wide-incident-angle chromatic polarized transmission
  on trilayer silver/dielectric nanowire gratings,'' \emph{Journal of the
  Optical Society of America B}, vol.~31, no.~5, pp. 1211--1216, May 2014.

\bibitem{Jung20TWC}
M.~{Jung} \emph{et~al.}, ``Performance analysis of large intelligent surfaces
  ({LIS}s): Asymptotic data rate and channel hardening effects,'' \emph{IEEE
  Transactions on Wireless Communications}, vol.~19, no.~3, pp. 2052--2065,
  Mar. 2020.

\bibitem{Wang20SPL}
P.~{Wang} \emph{et~al.}, ``Compressed channel estimation for intelligent
  reflecting surface-assisted millimeter wave systems,'' \emph{IEEE Signal
  Processing Letters}, vol.~27, pp. 905--909, 2020.

\bibitem{Wang20TWC}
Z.~{Wang} \emph{et~al.}, ``Channel estimation for intelligent reflecting
  surface assisted multiuser communications: Framework, algorithms, and
  analysis,'' \emph{IEEE Transactions on Wireless Communications}, vol.~19,
  no.~10, pp. 6607--6620, Oct. 2020.

\bibitem{Zheng20WCL}
B.~{Zheng} and R.~{Zhang}, ``Intelligent reflecting surface-enhanced {OFDM}:
  Channel estimation and reflection optimization,'' \emph{IEEE Wireless
  Communications Letters}, vol.~9, no.~4, pp. 518--522, Apr. 2020.

\bibitem{Wan20ICC}
Z.~{Wan} \emph{et~al.}, ``Broadband channel estimation for intelligent
  reflecting surface aided {mmWave} massive {MIMO} systems,'' in \emph{2020
  IEEE International Conference on Communications}, Dublin, Ireland, 2020, pp.
  1--6.

\bibitem{Matthiesen20WCL}
B.~{Matthiesen} \emph{et~al.}, ``Intelligent reflecting surface operation under
  predictable receiver mobility: A continuous time propagation model,''
  \emph{IEEE Wireless Communications Letters}, pp. 1--1, 2020.

\bibitem{Huang20arXiv}
\BIBentryALTinterwordspacing
Z.~{Huang} \emph{et~al.}, ``Transforming fading channel from fast to slow:
  {IRS}-assisted high-mobility communication.'' [Online]. Available:
  \url{https://arxiv.org/abs/2011.03147.}
\BIBentrySTDinterwordspacing

\bibitem{Rappaport02Book}
T.~S. Rappaport, \emph{Wireless Communications: Principles and Practice,
  \text{2nd ed}}.\hskip 1em plus 0.5em minus 0.4em\relax Upper Saddle River,
  NJ, USA: Prentice Hall, 2002.

\bibitem{Sun17SmallScale}
S.~{Sun} \emph{et~al.}, ``Millimeter wave small-scale spatial statistics in an
  urban microcell scenario,'' in \emph{2017 IEEE International Conference on
  Communications}, Paris, France, 2017, pp. 1--7.

\end{thebibliography}

\end{document}